\documentclass[aps,prc,twocolumn,twoside,showpacs,superscriptaddress]{revtex4}
\usepackage{graphicx}

\begin{document}

\newcommand{\tlab}{$T_{\mathrm{lab}}~$}
\newcommand{\plab}{$p_{\mathrm{lab}}~$}
\newcommand{\dsig}{$d\sigma/d\Omega~$}
\newcommand{\pppipi}{$\bar pp\to\pi^-\pi^+~$}

\title{Annihilation range and final-state interaction in $\bar{p}p$ annihilation into $\pi^-\pi^+$}
\author{B.~El-Bennich}
\email{bennich@physics.rutgers.edu}
\author{W.M.~Kloet}
\affiliation{Department of Physics and Astronomy, Rutgers University, \\ 
         136 Frelinghuysen Road, Piscataway, New Jersey 08854, USA}
\author{B.~Loiseau} 
\affiliation{Laboratoire de Physique Nucl\'eaire et de Hautes \'Energies (Groupe Th\'eorie), \\ 
         Universit\'{e} P.\& M. Curie, 4 Place Jussieu, 75252 Paris Cedex 05, France}
\date{29 January 2003}
\begin{abstract}
The large set of accurate data on differential cross section and analyzing power from the CERN LEAR experiment on 
\pppipi in the range from 360 to 1550 MeV/c is well reproduced within a distorted wave approximation approach. 
The initial $\bar pp$ scattering wave functions originate from a recent $\bar N N$ model. The transition operator 
is obtained from a combination of the $^{3}P_{0}$ and $^{3}S_{1}$ quark-antiquark annihilation mechanisms. A good 
fit to the data, in particular the reproduction of the double dip structure observed in the analyzing powers, 
requires quark wave functions for proton, antiproton, and pions with radii slightly larger than the respective 
measured charge radii. This corresponds to an increase in range of the annihilation mechanisms and 
consequently the amplitudes for total angular momentum $J=2$ and higher are much larger than in previous 
approaches. The final state $\pi\pi$ wave functions, parameterized in terms of  $\pi\pi$ phase shifts and 
inelasticities, are also a very important ingredient for the fine tuning of the fit to the observables.
\pacs{12.39.Jh.; 13.75.Cs; 21.30.Fe; 25.43.+t}
\end{abstract}
\maketitle 

\section{INTRODUCTION}

The very accurate set of data from the LEAR experiment \cite{hasan92} on \pppipi measuring the 
differential cross section and analyzing power from 360 to 1550 MeV/c is still a challenge for theoretical models 
after more than a decade. Large variations are observed in the analyzing power $A_{0n}$ as a function of angle at all 
energies, indicating presence of several partial waves already at low energies. 
However, recent model calculations \cite{moussallam83,mull91,mull92,bathas93,muhm96,yan96} lead to scattering amplitudes 
which are strongly dominated by total angular momentum $J=0$ and $J=1$. The reason for this is the choice of a rather 
short range annihilation mechanism. The short range of the annihilation in the model calculations originates from 
the dynamics of baryon exchange in Refs.~\cite{moussallam83,mull91,mull92,yan96} or from required overlap of 
quark and antiquark wave functions for proton and antiproton in Refs.~\cite{bathas93,muhm96,greenfinal}. 
On the other hand the experimental data on differential cross sections as well as those on asymmetries point to 
a significant $J=2$, $J=3$ and even higher $J$ contributions~\cite{oakden94,hasan94,kloet96,martin97}. 
 
All above mentioned models, for this reaction, use a distorted wave approximation (DWA).
The ingredients for calculating the \pppipi amplitudes consist of \textbf{i)} the initial 
$\bar pp$ scattering wave functions $\Psi_{\bar pp}({\mathbf r})$ \textbf{ii)} a transition operator 
$O({\mathbf r}',{\mathbf r})$ and \textbf{iii)} the final state $\pi\pi$ wave function $\Psi_{\pi\pi}({\mathbf r}')$.
The complete scattering amplitude {\bf {\em T}} itself, constructed in a DWA fashion, is
\begin{equation}
	\mbox{\bf{\em T}} =\int d{\mathbf r}'d{\mathbf  r}\ \Phi_{\pi\pi}({\mathbf r}') O({\mathbf r}',{\mathbf r})
	\Psi_{\bar pp}(\mathbf r).
	\label{1}
\end{equation}
For example in Ref.~\cite{bathas93} the transition operator $O({\mathbf r}',{\mathbf r})$ was obtained from a combination 
of $^{3}P_{0}$ and $^{3}S_{1}$ quark-antiquark annihilation model
\begin{equation}
 O({\mathbf r}',{\mathbf r})= N_0 [\,V_{^{3}\!P_{0}}({\mathbf r}',{\mathbf r})+\lambda\ V_{^{3}\!S_{1}}({\mathbf r}',{\mathbf r})\,] ,
\label{2}
\end{equation}
where the relative strength $\lambda$ is a complex parameter and $N_0$ an overall real normalization factor. In the same 
reference $\Psi_{\bar pp}({\mathbf r})$ was provided by the 1982 Paris $\bar NN$ potential model~\cite{cote82} and 
$\Phi_{\pi\pi}({\mathbf r}')$ was a simple plane wave. This work did not succeed in reproducing the double-dip structure 
of the analyzing power and the forward peak in the differential cross section as seen experimentally at, for example, 497 MeV/c. 
All previously mentioned models exhibit similar difficulties. 

The aim of the present paper is to study possible improvements of previous models. First of all, mesonic final-state 
interaction should be considered. The total energy of the $\bar pp\to 2\pi$ reaction for the 
studied data set is in the 2 GeV range. In this energy region the $\pi\pi$ interaction is characterized 
by several resonances \cite{pdg02}. In Refs.~\cite{mull92,muhm96} the role of $\pi\pi$ final-state interactions was studied.
In Ref.~\cite{mull92} some improvement was obtained using a $\pi\pi$ model reproducing the real part of the $\pi\pi$ phase shifts with 
inelasticity parameters in all $J \neq 0$ partial waves remaining close to 1. In Ref.~\cite{muhm96} which explores the $^3P_0$ 
part of the quark-antiquark dynamics in the transition operator, the final-state interaction, resulting from meson exchange, affects mainly 
observables in the backward region. In both approaches the double-dip structure observed in experimental data of the analyzing power remains 
elusive and further study is still needed. Final-state interactions of two mesons in $N\bar N$ annihilation have also been studied, at quark level, within 
an extension of the quark rearrangement model \cite{greenfinal}. 
Results were reported for branching ratios of decays into various two-meson channels. Unfortunately 
there are no predictions from this work for differential cross sections or analyzing powers.

Within the approach of Ref.~\cite{bathas93} we will study the effect of final-state interactions 
guided by the $\pi\pi$ coupled channel model of Ref.~\cite{kloet98}. Secondly we will study predictions following modification of the 
annihilation operator $O({\mathbf r}',{\mathbf r})$. As remarked above this operator is rather short range in all present models.
The reason for the short range of $O({\mathbf r}',{\mathbf r})$ is that in the quark-antiquark annihilation model the antiproton
and proton have a relatively small radius since their quark wave functions describe only the $qqq$ and $\bar q\bar q\bar q$
core ignoring the $\bar qq$ cloud. It could be a cause of discrepancy between theory and experiment. In Ref.~\cite{bathas93} 
it has already been noticed that an increase of the annihilation range improves substantially the theoretical description of 
the data.

In the present work it is shown that the $\pi \pi$ final-state interaction is a very significant tool for the fine tuning of the fit 
to the observables. Furthermore the parameters that determine the size of protons and pions are also crucial. An increase of both 
proton and pion sizes, in closer agreement with their measured radii, allows for a much better fit to the experimental cross sections 
and analyzing powers. Expressions of the observables in term of the basic amplitudes together with the DWA ingredients are briefly 
recalled in Section II. The description of the final-state interaction is performed in Section III. The modifications of the 
range of the  annihilation mechanisms are studied in Section IV. Section V presents the final results and conclusions are summarized 
in Section VI.

\section{OBSERVABLES AND DWA INGREDIENTS}

The reaction $\bar pp\to \pi^-\pi^+$ can be fully described in the helicity formalism by two independent helicity amplitudes 
$F_{++}(\theta)$ and $F_{+-}(\theta)$. The angle $\theta$ is the c.m. angle between the outgoing $\pi^-$ and the incoming $\bar p$.
There are four possible observables \cite{martin80}
\begin{equation}
\label{3 }
\frac{d\sigma}{d\Omega}=\frac{1}{2}(\vert F_{++}\vert^2+\vert F_{+-}\vert^2),
\end{equation}
\begin{equation}
\label{4}
A_{0n} \frac{d\sigma}{d\Omega}=\mathrm{Im}\,(F_{++}F_{+-}^*),
\end{equation}
\begin{equation}
\label{5}
A_{\ell s} \frac{d\sigma}{d\Omega}=\mathrm{Re}\,(F_{++}F_{+-}^*),
\end{equation}
\begin{equation}
\label{6}
A_{ss} \frac{d\sigma}{d\Omega}=\frac{1}{2}(\vert F_{++}\vert^2-\vert F_{+-}\vert^2).
\end{equation}
So far only \dsig and $A_{0n}$ have been accurately measured at 
LEAR~\cite{hasan92} .
For completeness we recall here the partial wave expansion of the helicity amplitudes \cite{martin80}
\begin{eqnarray}
\label{eq:7}
F_{++}(\theta)  =  \frac{1}{p}\sum_J\sqrt{(2J+1)/2} \nonumber \hspace{2.3cm} \\
 \times\!\left[\sqrt{J}f_{J-1}^J-\sqrt{J+1}f_{J+1}^J\right ]\!\times\!P_J(\cos\theta),
\end{eqnarray}
and 
\begin{eqnarray}
\label{eq:8}
F_{+-}(\theta) = \frac{1}{p}\sum_J\sqrt{(2J+1)/2} \nonumber \hspace{2.3cm} \\
\times\!\left[\sqrt{\frac{1}{J}}f_{J-1}^J+\sqrt{\frac{1}{J+1}}f_{J+1}^J \right ]\!\times\!P^\prime_J(\cos\theta).
\end{eqnarray}
The indices $J$ and $L=J\pm1$ are the total and orbital angular momentum of the $\bar pp$ system respectively.
$P_J(\cos\theta)$ and $P^\prime_J(\cos\theta)$ denote a Legendre polynomial and its derivative. The angular momentum 
of the $\pi\pi$ system is $\ell_{\pi\pi}\equiv J$. Because of parity conservation there are no $L=J$ amplitudes in above 
expansion. Total isospin $I=0$ for even $J$ and $I=1$ for odd $J$. In Eqs. (\ref{eq:7}) and (\ref{eq:8}) $p$ is the magnitude 
of the antiproton center of mass (CM) momentum.

The partial wave amplitudes $f_L^J$ for $L=J\pm 1$ are calculated following the DWA method of Eq.~(\ref{1}).
One ingredient is the initial coupled spin-triplet $\Psi_{\bar pp}(\mathbf {r})$ wave function in configuration 
space as obtained in \cite{elbennich98}.
The operator $O(\mathbf{r},\mathbf{r'})$ is constructed from the quark model description of protons and pions combined with the 
$^3P_0$ and $^3S_1$ quark-antiquark annihilation and rearrangement mechanism \cite{bathas93}. The last ingredient is the $\pi\pi$ 
scattering wave function $\Phi_{\pi\pi}(\mathbf{r'})$ which in this paper will be built according to a study of a realistic 
$\pi\pi$ scattering model~\cite{kloet98}, while also comparisons will be made for a simple plane wave $\pi\pi$ final state. 
Subsequently one obtains the differential cross section \dsig and the analyzing power $A_{0n}$ or left-right asymmetry for the 
proton target polarized normal to the scattering plane.

\begin{figure*}
\begin{center}
\includegraphics[angle=90,scale=0.6]{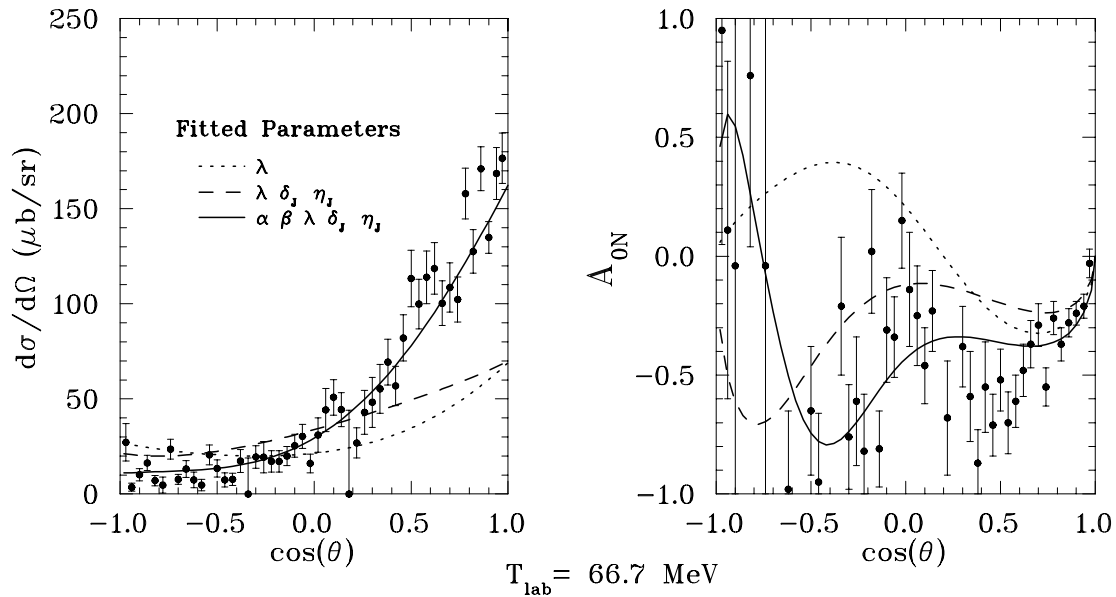}
\end{center}
\caption{Differential cross section and analyzing power of the reaction \pppipi at \tlab= 66.7 MeV (\plab= 360 MeV/c).
         Experimental data from Hasan {\em et al}. \cite{hasan92}. The different curves are described in the text.}
\label{fig1}
\end{figure*}
\begin{figure*}
\begin{center}
\includegraphics[angle=90,scale=0.6]{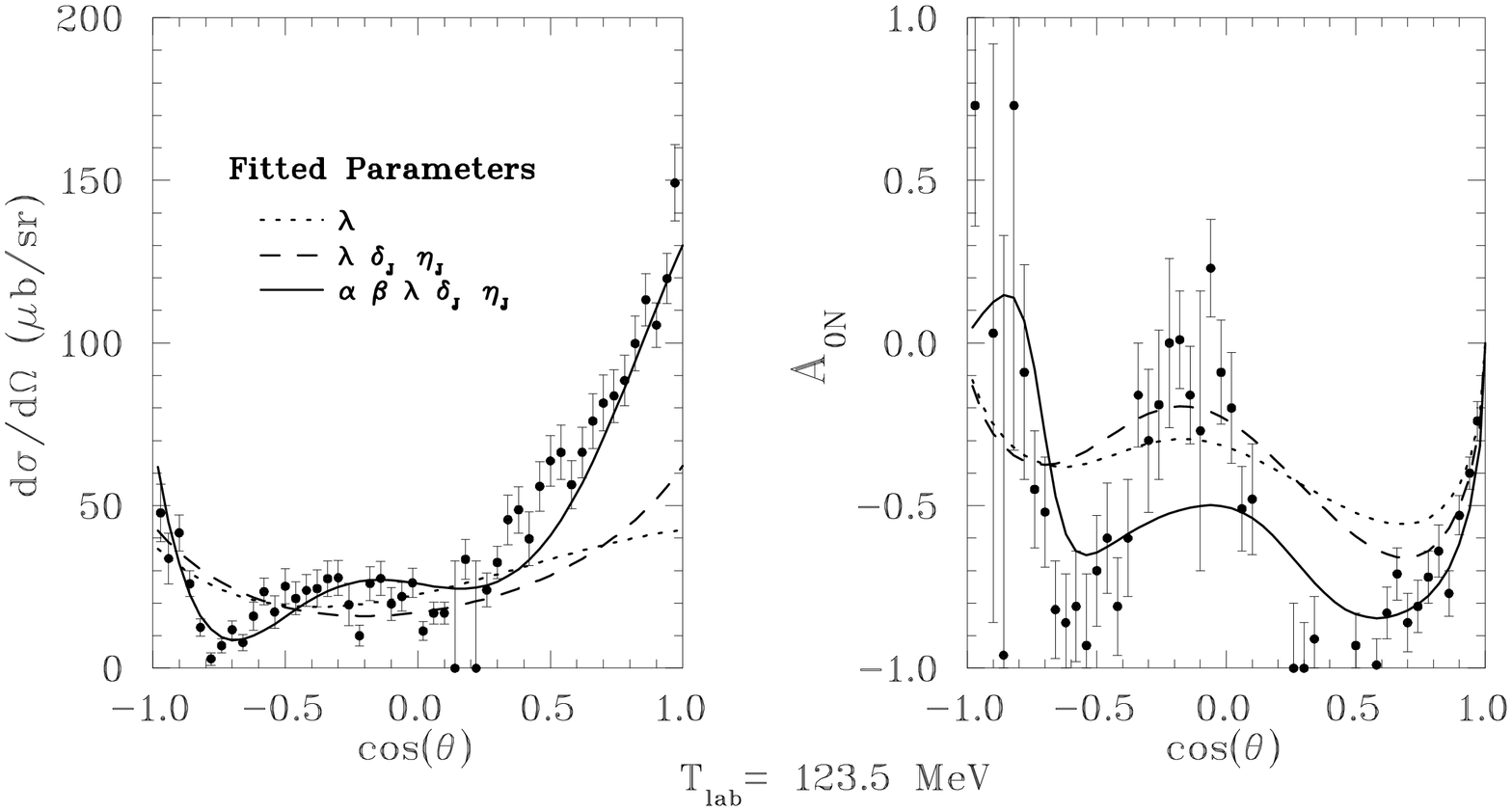}
\end{center}
\caption{As in Fig.~\ref{fig1} but for \tlab= 123.5 MeV  (\plab = 497 MeV/c).}
\label{fig2}
\end{figure*}
\begin{figure*}
\begin{center}
\includegraphics[angle=90,scale=0.6]{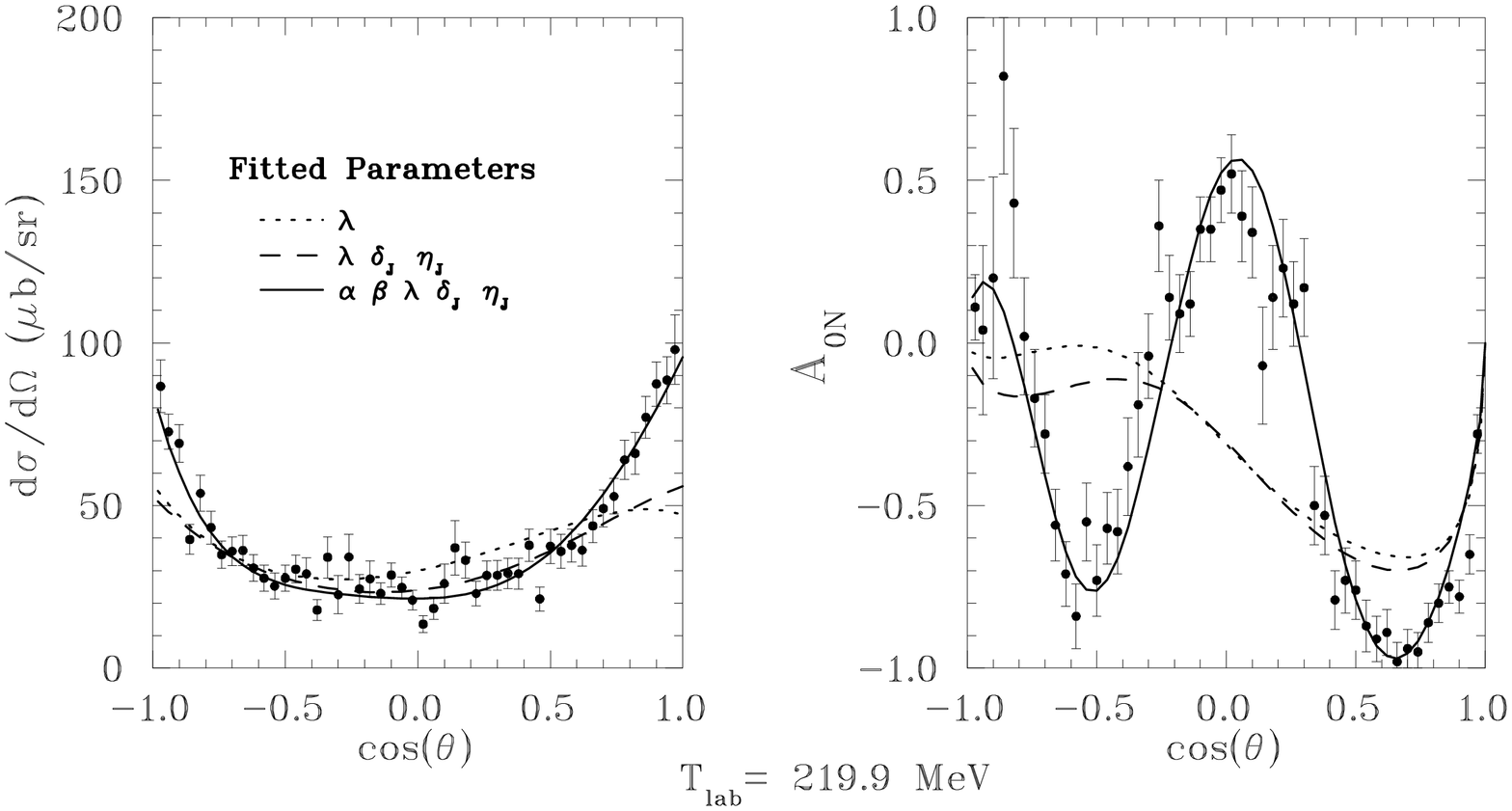}
\end{center}
\caption{As in Fig.~\ref{fig1} but for  \tlab= 219.9 MeV (\plab= 679 MeV/c).}
\label{fig3}
\end{figure*}
\begin{figure*}
\begin{center}
\includegraphics[angle=90,scale=0.6]{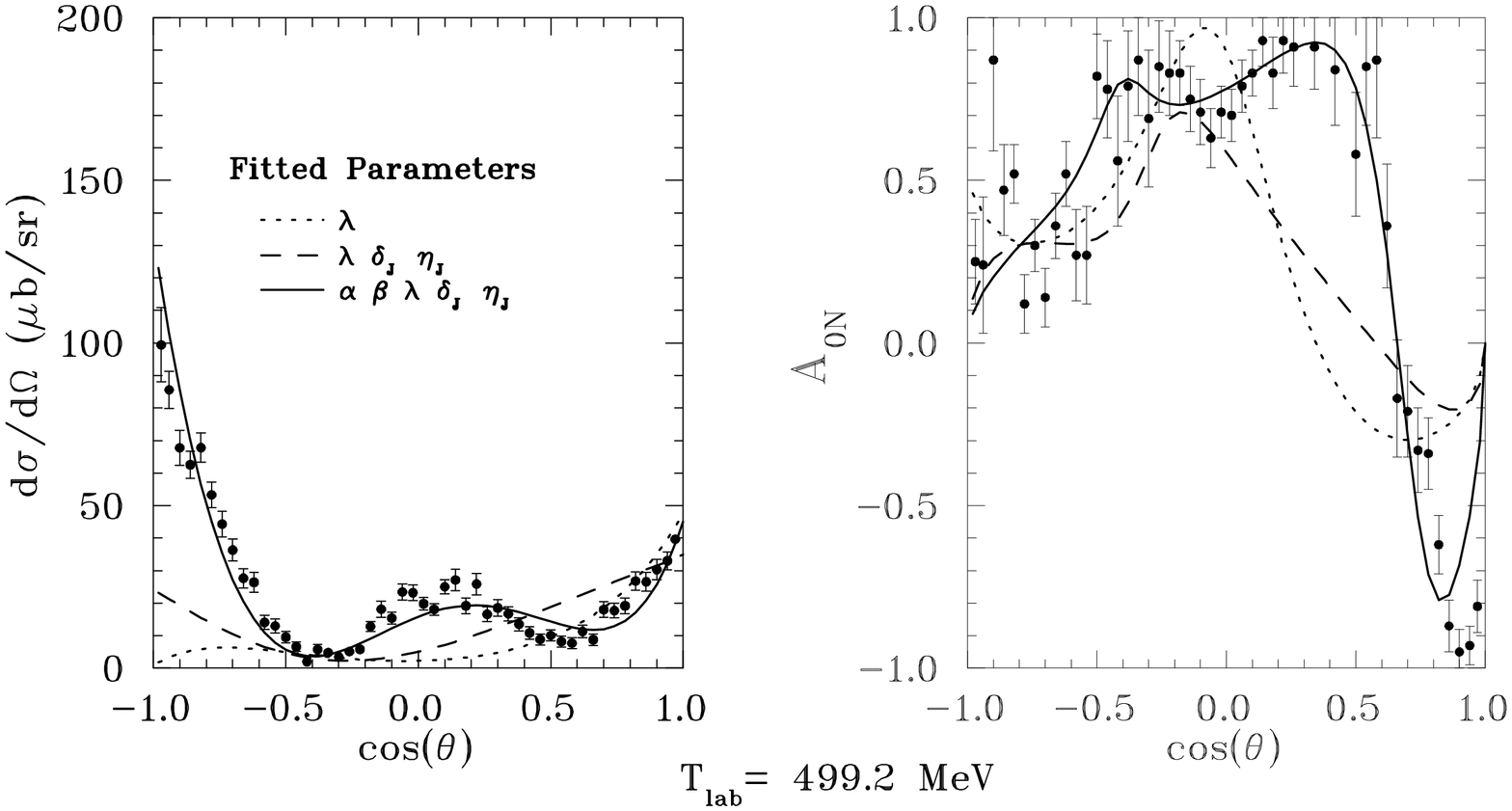}
\end{center}
\caption{As in Fig.~\ref{fig1} but for \tlab= 499.2 MeV (\plab= 1089 MeV/c).}
\label{fig4}
\end{figure*}

\section{FINAL-STATE INTERACTION\label{sec3}}

\begin{figure*}
\begin{center}
\includegraphics[angle=90,scale=0.6]{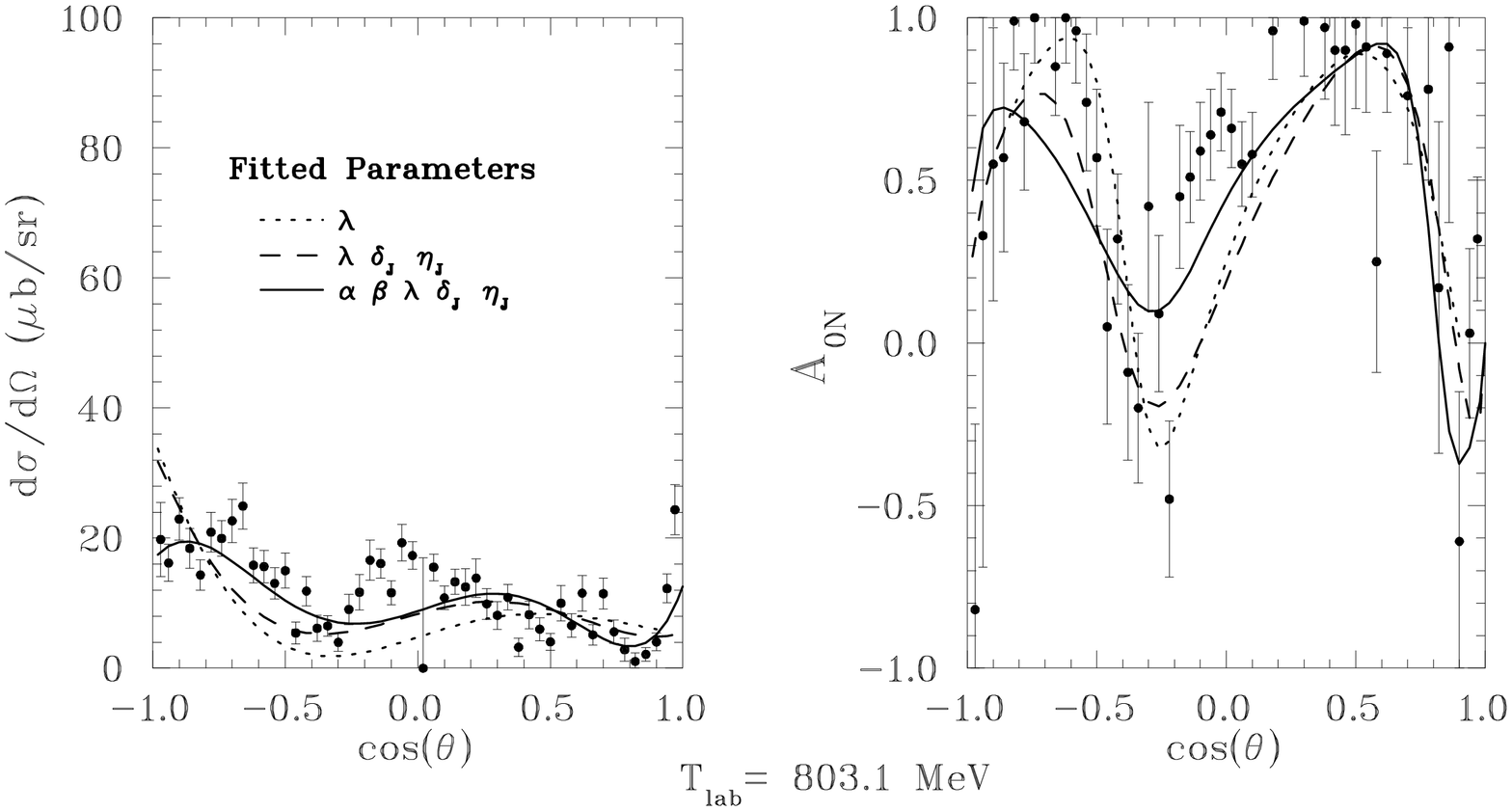}
\end{center}
\caption{As in Fig.~\ref{1} but for \tlab= 803.1 MeV  (\plab= 1467 MeV/c).}
\label{fig5}
\end{figure*}

First we assume no interaction between the final pions and describe the $\pi\pi$ scattering wave function $\Phi_{\pi\pi}(\mathbf{r'})$ 
as a plane wave. In this case the overall normalization of \dsig and the relative strength $\lambda$ between $^3P_0$ and $^3S_1$ 
annihilation mechanism are parameters to be determined through $\chi^2$ minimization. Out of the twenty energies where polarization data are 
available~\cite{hasan92} we chose a representative set of five energies, \tlab= 66.7, 123.5, 219.9, 499.2, and 803.1 MeV, corresponding 
to antiproton momenta of respectively \plab= 360, 497, 679, 1089, and 1467 MeV/c. \tlab is the laboratory kinetic energy of the 
antiproton beam. At 123.5 MeV (\plab= 497 MeV/c) and 219.9 MeV (\plab= 679 MeV/c) we do reproduce the results of~\cite{bathas93} using 
their parameters. Results for the set of five energies are shown in Figs.~\ref{fig1}--\ref{fig5} as short-dashed curves.
The overall fits are poor with exception of the analyzing power $A_{0n}$ at 803.1 MeV (\plab= 1467 MeV/c). The double-dip structure of 
$A_{0n}$ is not reproduced at the three lowest energies of the set considered here, which can be attributed to the very small values of the 
$J\geq2$ amplitudes predicted by this model. A lack of substantial $J\geq2$ amplitudes is also evident in the predictions for \dsig 
at all energies. The forward peak is poorly reproduced, and also the backward peak, prominent in the data at \tlab= 499.2 MeV, 
is missing in this simplified model. Similar findings were obtained previously at \tlab= 123.5 and 219.9 MeV by ~\cite{bathas93}.
As shown in the analysis of~\cite{kloet96} we recall that large $J\geq 2$ amplitudes are needed to explain the angular dependence of 
$A_{0n}$ and \dsig. 

One possibility to enhance the amplitudes of higher $J$ values could be to introduce $\pi\pi$ final-state interaction.
The elastic $\pi\pi\to \pi\pi$ amplitude is known from threshold up to the total relativistic $\pi\pi$ energy $\sqrt{s}= 1800$ MeV mainly 
from analysis of the $\pi N\to \pi\pi N$ reaction. The extracted $\pi\pi\to \pi\pi$ amplitudes can be parameterized in terms of phase shifts
$\delta_J$ and inelasticities $\eta_J$ where $J=$ 0, 1, 2 and 3 \cite{bunch}. In~\cite{kloet98} a coupled channel model of $\pi\pi,\ \bar KK$ 
and $\rho\rho$ was proposed to reproduce these phase parameters $\delta_J$  and $\eta_J$ for $J=$ 0, 1, 2 and 3.
The $\Phi_{\pi\pi}(\mathbf{r'})$ wave functions required for the final-state interaction of $\bar pp\to \pi\pi$, can be constructed from this 
coupled channel model in a straightforward manner. However, since the needed energy range in $\bar pp\to \pi\pi$ is from $\sqrt{s}=1910$ to 
2272 MeV one has to rely on extrapolation of the coupled channel model results beyond $\sqrt{s}=1800$ MeV. 
Calculations with the corresponding $\Phi_{\pi\pi}(\mathbf{r'})$ show a significant sensitivity of the observables to the $\pi\pi$ final-
state interaction. Even so the $\pi\pi$ scattering amplitude from the extrapolated coupled channel $\pi\pi$ model still does not improve 
the predictions of $\bar pp\to \pi\pi$ obtained previously with $\pi\pi$ plane waves. But in this study we did find that the off-shell part 
of $\Phi_{\pi\pi}(\mathbf{r'})$ does play a very minor role. In particular it was observed that the same predictions of observables 
\dsig and $A_{0n}$ can be obtained using only the asymptotic part of the $\pi\pi$ wave functions. We then exploit this fact by 
asserting for the remainder of this paper that in $\bar pp\to \pi\pi$ the unknown $\pi\pi$ final-state scattering can be fully described by 
the asymptotic $\pi\pi$ wave functions parameterized by the $\pi\pi$ phases $\delta_J$ and inelasticities $\eta_J$. 
\begin{figure*}
\begin{center}
\includegraphics[angle=90,scale=0.6]{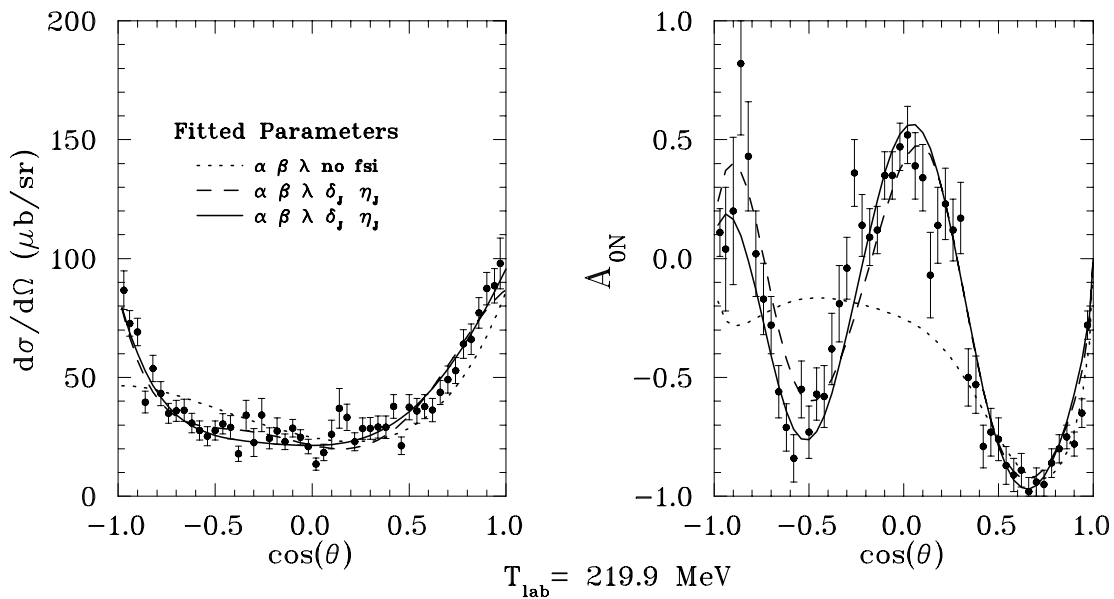}
\end{center}
\caption{Various fits for \tlab= 219.9 MeV  (\plab= 679 MeV/c). 
The solid curve is as in Fig.~\ref{fig3}.
The long-dashed curve represents a fit with final-state interaction, where $\alpha=1.09$ fm$^{-2}$ and $\beta=1.51$ fm$^{-2}$. 
The short-dashed curve is a fit with no final-state interaction, where $\alpha=1.27$ fm$^{-2}$ and $\beta=1.53$ fm$^{-2}$. 
}
\label{fig3b}
\end{figure*}

We then avoid the problem of extrapolation to high $\sqrt{s}$ by including these new parameters $\delta_J$ and $\eta_J$ in the minimization 
process to obtain realistic fits to \dsig and $A_{0n}$. Results of this second fitting procedure are shown as the long-dashed 
curves in Figs.~\ref{fig1}--\ref{fig5}. Switching on final state $\pi\pi$ interaction improves the fit of $A_{0n}$ by 
readjusting the strength of the helicity amplitudes of different $J$. This allows for prediction of a double-dip structure at lower 
energies. This feature is a crucial requirement of the data and shows the need for incorporating the final-state interaction of the pions 
in some form. However the predictions for \dsig show only a modest improvement over the model without final-state interaction and the 
question arises whether there is additional freedom within the model to ameliorate the present fit. So far, the only variable parameter 
in the annihilation operator is the relative strength $\lambda$ unless one allows variations of the range of the annihilation mechanism 
controlled by the parameters $\alpha$ and $\beta$ \cite{bathas93}. In the next section we will investigate the effects of variations in 
$\alpha$ and $\beta$.

\section{MODIFICATION OF THE ANNIHILATION RANGE}

In order to derive the annihilation operator $O(\mathbf{r'},\mathbf{r})$ (\ref{2}), one can describe the proton, antiproton and pions in 
terms of quarks and antiquarks with the use of Gaussian wave functions \cite{bathas93}. 
This amounts to approximate quark confinement by solving the Dirac equation with either a scalar or vector harmonic oscillator potential.
The proton (antiproton) intrinsic wave function for the annihilation mechanism is :
\begin{eqnarray}
\label{9}
\psi_{_N}({\mathbf r}_1,{\mathbf r}_2, {\mathbf r}_3)= N_{\!_N} \exp \left[ -\frac{\alpha}{2}\sum_{i=1}^3({\mathbf r}_i-{\mathbf r}_{\!_N})^2
\right] \nonumber \\
\times \chi_{_N}\mathrm{(spin, isospin, color)},
\end{eqnarray}
where  ${\mathbf r}_i$ are the quark (antiquark) coordinates and ${\mathbf r}_{\!_N}$ the nucleon (antinucleon) coordinate.
An $S$-wave meson intrinsic wave function is:
\begin{eqnarray}
\label{10}
\phi_{_M}({\mathbf r}_1, {\mathbf r}_4)=N_{\!_M} \exp \left[ -\frac{\beta}{2}\sum_{i=1,4}( {\mathbf r}_i-{\mathbf r}_{_M})^2 \right] 
 \nonumber \\
\times \chi_{_M} \mathrm{(spin, isospin, color).}
\end{eqnarray}
Here ${\mathbf r}_1$ and $ {\mathbf r}_4$ are the quark and antiquark coordinates, respectively. The coordinate of the meson 
is ${\mathbf r}_{_M}$.

Typical parameter values used before are $\alpha=2.80$~fm$^{-2}$ and  $\beta=3.23$~fm$^{-2}$ 
which correspond to a proton (antiproton) radius of 0.60~fm and a meson radius of 0.48~fm \cite{green}. This value of $\alpha$ describes the 
$qqq$ ($\bar q \bar q \bar q$) core of the proton (antiproton) while the measured charge radius, which for the proton is about 0.8~fm, 
includes also the mesonic cloud.
Explicit expressions in terms of $\alpha$ and $\beta$ for the transition operators $V_{^3P_0}({\mathbf r'}, {\mathbf r})$ and 
$V_{^3S_1}( {\mathbf r'}, {\mathbf r})$ of Eq.~(\ref{2}) can be found in \cite{bathas93}.
However, one can also argue that in modeling the Gaussian wave functions as in Eqs. (\ref{9}) and (\ref{10}), values for $\alpha$ 
and $\beta$ simply should be in accordance with known charge radii of the proton and pion. The measured pion charge distribution 
radius \cite{amendolia} is $\langle r^2_{\pi}\rangle^{\!1/2}= 0.663\pm0.006$~fm. 
For the proton we find $\langle r^2_{p}\rangle^{1/2}= 0.870\pm0.008$~fm in the literature \cite{pdg02}.
We are thus left with a certain freedom when it comes to choosing values for $\alpha$ and $\beta$ and we can wonder about the effects 
on the annihilation mechanism. The values of the parameters $\alpha$ and $\beta$ determine effectively the range of the annihilation 
mechanism. Let us assume we increase the size of the proton and the pion from their original values in \cite{bathas93} by decreasing 
$\alpha$ and $\beta$. Then the integration over the intrinsic quark coordinates of, for example, the wave functions 
$\psi_p({\mathbf r}_1,{\mathbf r}_2, {\mathbf r}_3)$,  
$\psi_{\bar{p}}( {\mathbf r}_4,{\mathbf r}_5, {\mathbf r}_6)$, 
$\phi_{\pi^+}({\mathbf r}_1, {\mathbf r}_4)$, and 
$\phi_{\pi^-}({\mathbf r}_2, {\mathbf r}_5)$ 
will result in a larger quark overlap. The corresponding annihilation operator $O({\mathbf r}',{\mathbf r})$ will have a longer range 
and therefore higher partial waves will contribute to the total amplitude {\bf {\em T}} in Eq.~(\ref{1}) as required by the analyses of 
experimental data of Refs.~\cite{oakden94,hasan94,kloet96,martin97}.

Changes of $\alpha$ and $\beta$ can furthermore be linked to relativistic corrections of the pion wave function 
$\phi_M({\mathbf r}_1,{\mathbf r}_4)$ as the outgoing energy of the pion is much larger than its rest mass. In practical terms this means 
that in the center of mass, where the calculation is performed, the pion wave function should not be described as a sphere anymore. 
A proper treatment requires a Lorentz boost of the pion intrinsic wave function from its rest frame to the CM frame. Thus one expects 
a change in the geometry of the Gaussian shape of the pion \cite{elbennich02}. This affects the overlap integral (\ref{1}) which could 
be mocked by a simultaneous alteration of $\alpha$ and $\beta$. We therefore take $\alpha$ and $\beta$  to be variable parameters to 
the extent that they still satisfy physical conditions.

\begin{figure*}
\begin{center}
\includegraphics[angle=90,scale=0.6]{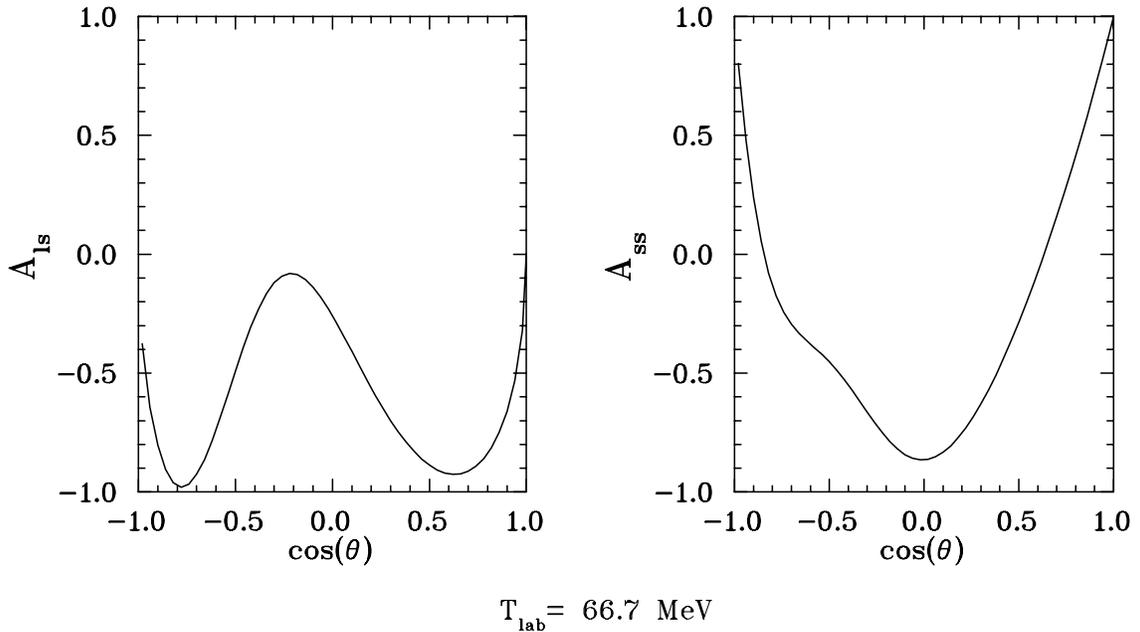}
\end{center}
\caption{Predictions at \tlab= 66.7 MeV  (\plab= 360 MeV/c) for the observables $A_{\ell s}$ and $A_{ss}$ obtained with the final fit.}
\label{fig6}
\end{figure*}

\section{FINAL FIT}

The results of our final fit which includes now values of the parameters 
$\alpha, \ \beta,\ \lambda=|\lambda| \exp(i\theta_{\lambda}),\ \delta_J,\ \eta_J,\ $
and of the overall normalization $N_0$ are shown as solid curves in Figs.~\ref{fig1}--\ref{fig5}. 
The improvement over the previous fits is dramatic but requires both increased sizes of the antiproton, proton and pions as 
well as a tuned final-state interaction. The main achievement is the reproduction in the differential cross section \dsig 
of the characteristic forward peaks at lower energies (\tlab= 66.7, 123.5 and 219.9 MeV) and backward peaks at higher energies 
(\tlab= 219.9 and 499.2 MeV). This clearly indicates that the increase of the annihilation range now produces much larger 
amplitudes for $J=2$ and higher. The predictions for the double-dip structure of the analyzing power $A_{0n}$ compare much 
better with the experimental results especially for \tlab$\ge 219.9$ MeV.

In addition to the double-dip structure of $A_{0n}$ the experimental data display another characteristic feature: the asymmetry 
shifts from predominantly negative values at lower energies toward positive values at higher energies. Our final fit 
accounts for this pattern. The quark model parameters $\alpha$, $\beta$, $\lambda$, and the overall norm $N_0$ resulting from this fit 
are listed in Table~\ref{model}  and the phase shifts $\delta_J$ and inelasticities $\eta_J$ of the final-state interaction with their dependence 
on $\sqrt s$ in Table~\ref{phases}. Note that the latter have been readjusted in the final fit and differ from the $\delta_J$ and $\eta_J$ 
obtained in the fit presented in section \ref{sec3}. The $\pi \pi$ phases for $J =$ 0, 1 are small but for $J = $ 2, 3, 4 they are substantial. 
The inelasticities $\eta_J$, in particular for $J = $~2 may indicate the presence of resonances in this channel. From Table~\ref{model} 
it is also clear that the parameters $\alpha$ and $\beta$ are almost constant with energy  but have lower values at 219.9~MeV. 
 Nevertheless, in Fig.~\ref{fig3b} we show (long-dashed line) that one can obtain a fit of similar quality keeping the size parameters
$\alpha=1.09$~fm$^{-2}$ and $\beta=1.51$~fm$^{-2}$ close to the ones at the other four energies.
The relative strength $\lambda$ exhibits an energy dependence. This dependence shows a smooth decrease of $|\lambda|$ with increasing $\sqrt s$, 
which indicates that the $^3P_0$ mechanism preponderates at higher energies. The phase $\theta_{\lambda}$ of $\lambda$ should be compared with 
the value $\theta_{\lambda}=180^{\circ}$ of ~\cite{bathas93}. Furthermore, we find that $\alpha$ decreases by about a factor 2.3 while 
$\beta$ decreases by about 2.2. In other words, the proton size in this model increases to $\langle r^2_{p}\rangle^{\!1/2}=0.91$~fm and the pion 
radius is now $\langle r^2_{\pi}\rangle^{\!1/2}=0.71$~fm, which is within 7\% of the values of references \cite{pdg02} and \cite{amendolia} 
mentioned in the previous section. 

\begin{table}[b]
\begin{ruledtabular}
  \begin{tabular}{c|lllll}   
    \tlab [MeV]  & 66.7  & 123.5 & 219.9 & 499.2 & 803.1 \\
    \plab [MeV/c] & 369 & 497 & 679 & 1089 & 1467  \\ \hline
     $\alpha$ [fm$^{\!-2}$] & 1.20 & 1.19 & 1.11 & 1.20 & 1.20 \\
     $\beta$ [fm$^{\!-2}]$ & 1.54 & 1.51 & 1.00 & 1.54 & 1.54 \\
     $|\lambda|$ & 1.200 & 1.064 & 0.662 & 0.545 & 0.424 \\
     $\theta_{\lambda} [^{\circ}]$ & 197.10 & 125.68 & 166.83 & 176.32 & 183.29 \\ 
     $N_0$ $[10^5]$ & 5.251 & 4.804 &  5.046 &  4.222 & 4.667 
  \end{tabular}
\end{ruledtabular}
\caption{Quark model parameters as function of \tlab.} 
\label{model}
\end{table}

\begin{table}[b]
\begin{ruledtabular}
  \begin{tabular}{c|ccccc} 
    \tlab [MeV]  &  66.7  & 123.5 & 219.9 & 499.2 & 803.1 \\
    $\sqrt s$ [MeV] & 1910. & 1937. & 1983. & 2111. & 2242. \\ \hline
    $\eta_0$ &  0.840 & 0.989 & 0.937 & 1.00 & 1.00 \\
    $\delta_0$ & -4.57 & -4.37 & -6.59 & -5.24 & -0.63 \\ 
    $\eta_1$ & 0.998 & 0.995 & 0.921 & 0.999 & 0.914 \\
    $\delta_1$ & 8.05 & 4.89 & -4.14 & -3.24 & -1.88 \\
    $\eta_2$ & 0.756 & 0.669 & 0.760 & 0.818 & 0.309 \\
    $\delta_2$ & 35.74 & 24.49 & 55.57 & 37.66 & 74.95 \\ 
    $\eta_3$ & 1.00 & 1.00 & 0.808 & 1.00 & 1.00  \\ 
    $\delta_3$ & 59.66 & 41.41 & 62.45 & 48.15 & -19.59 \\
    $\eta_4$ & 1.00 & 0.997 & 0.858 & 1.00 & 0.990 \\
    $\delta_4$ & 36.93 & -41.29 & 6.19 & -60.70  & 61.77 \\ 
  \end{tabular}
\end{ruledtabular}
  \caption{Phases shifts $\delta_J$ and inelasticities $\eta_J$ of the final-state interaction for $0\le J\le 4$ 
           as function of $\sqrt s$.}
\label{phases}
\end{table}

\begin{figure*}
\begin{center}
\includegraphics[angle=90,scale=0.6]{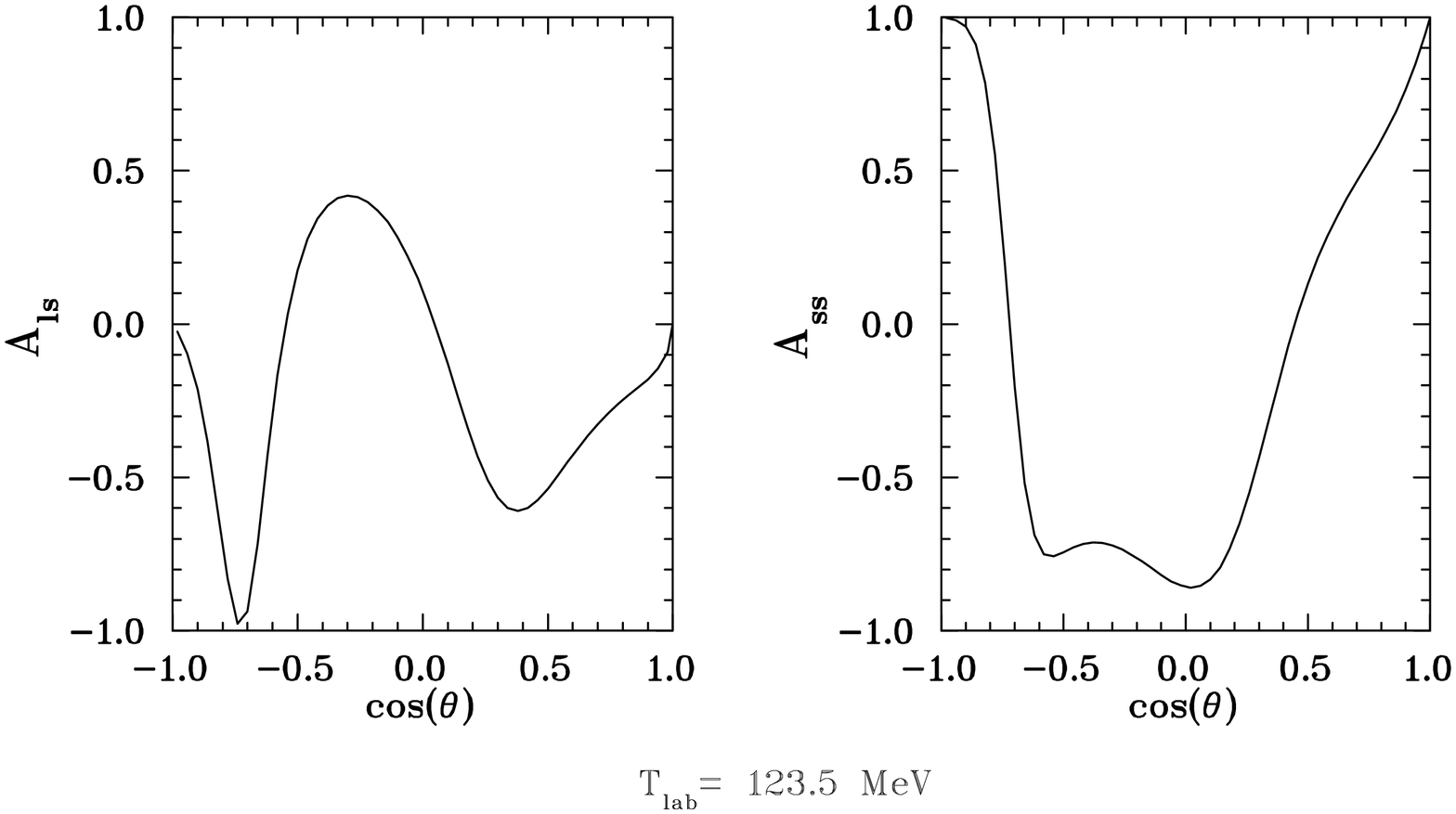}
\end{center}
\caption{As in Fig.~\ref{fig6} but for \tlab= 123.5 MeV  (\plab= 497 MeV/c).}
\label{fig7}
\end{figure*}
\begin{figure*}
\begin{center}
\includegraphics[angle=90,scale=0.6]{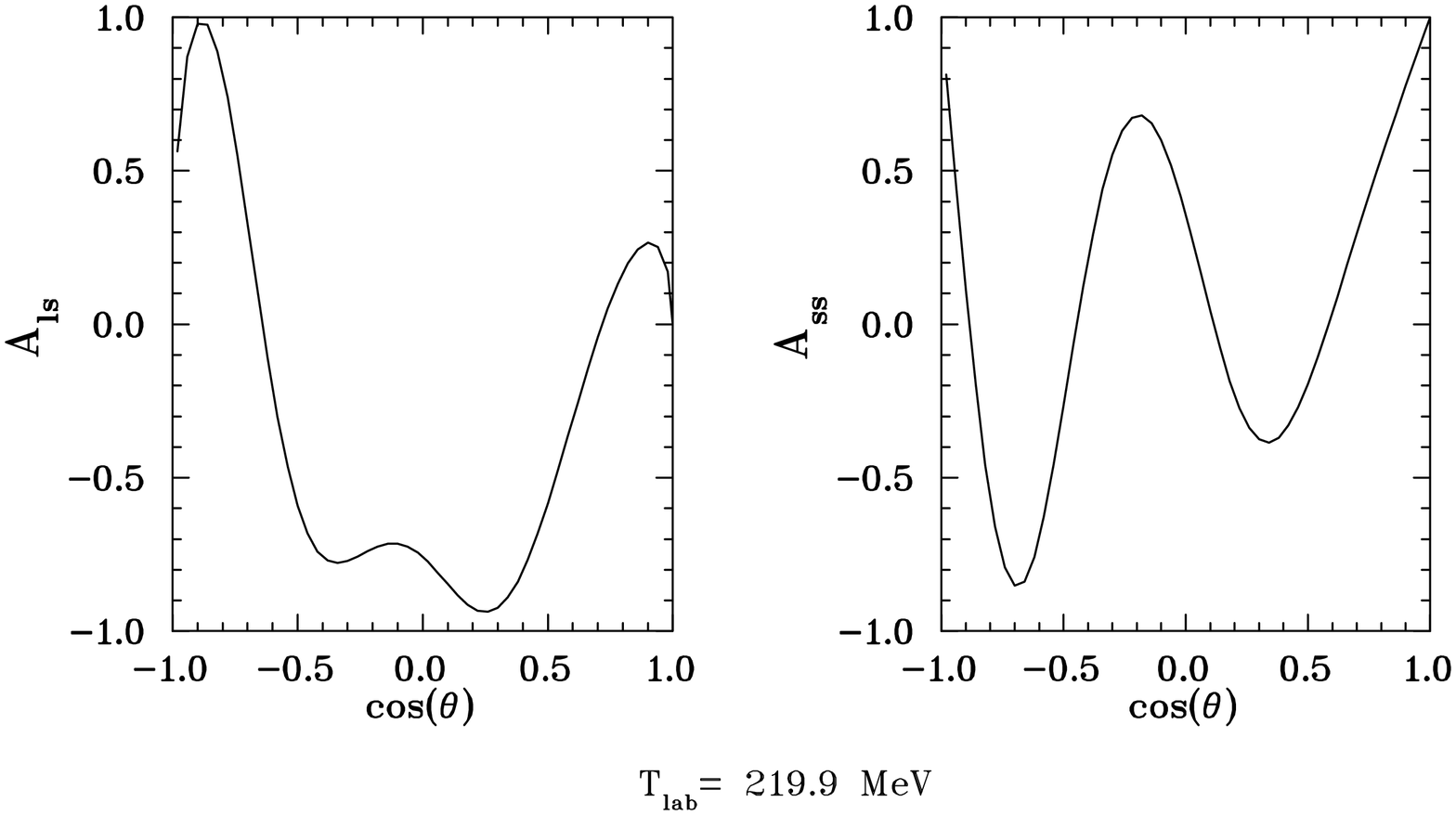}
\end{center}
\caption{As in Fig.~\ref{fig6} but for \tlab= 219.9 MeV  (\plab= 679 MeV/c).}
\label{fig8}
\end{figure*}
\begin{figure*}
\begin{center}
\includegraphics[angle=90,scale=0.6]{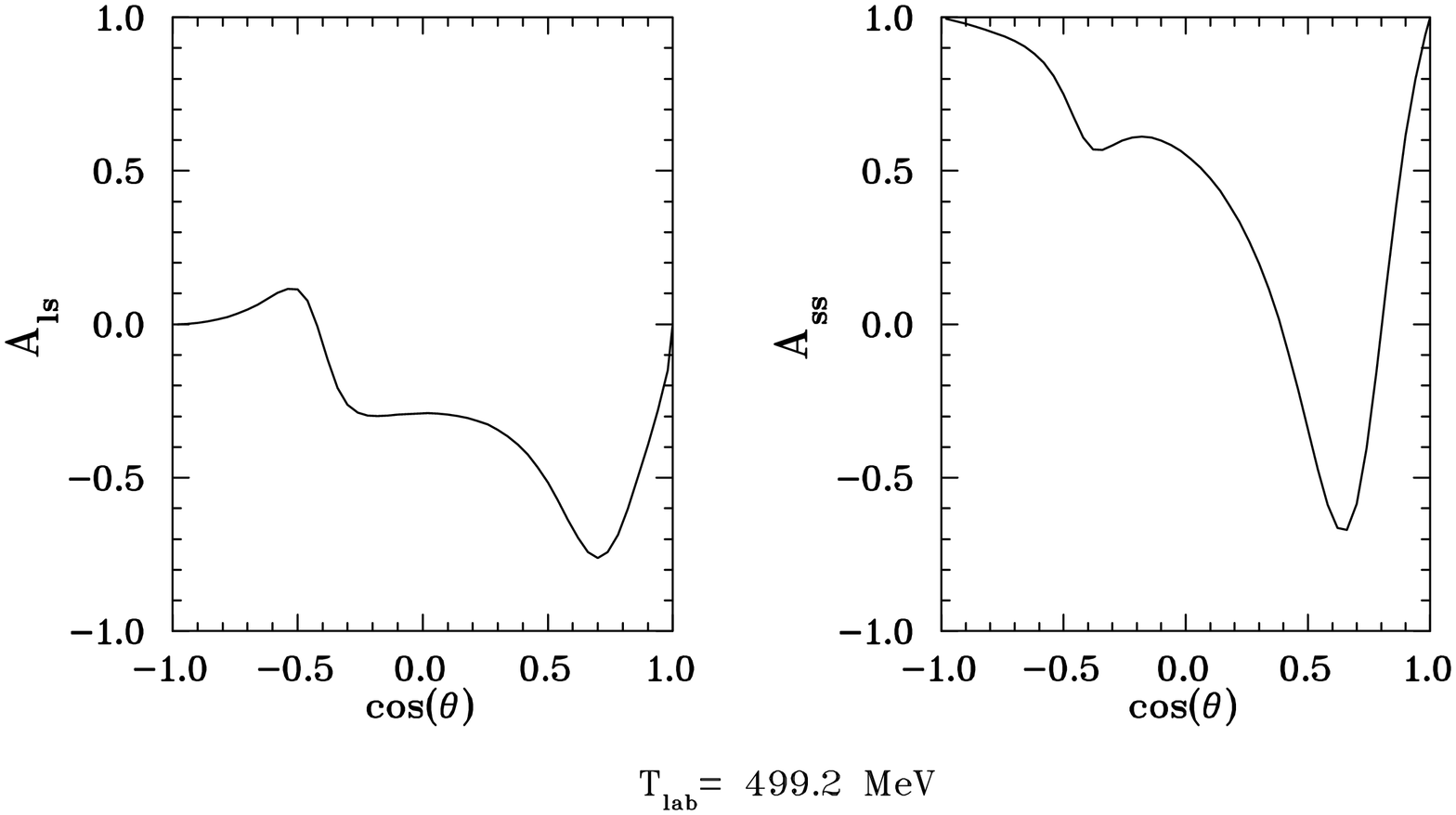}
\end{center}
\caption{As in Fig.~\ref{fig6} but for \tlab= 499.2 MeV  (\plab= 1089 MeV/c).}
\label{fig9}
\end{figure*}
\begin{figure*}
\begin{center}
\includegraphics[angle=90,scale=0.6]{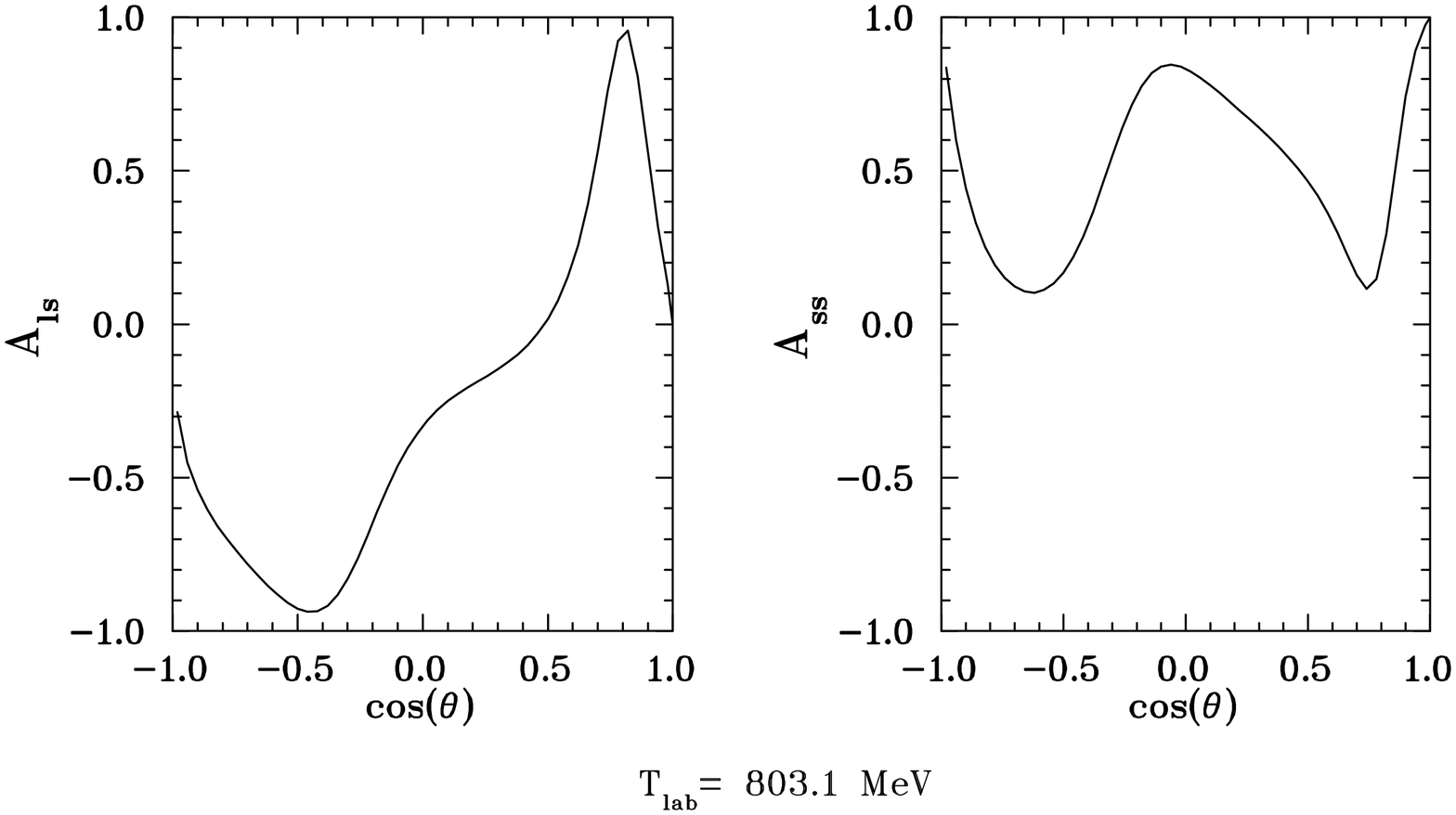}
\end{center}
\caption{As in Fig.~\ref{fig6} but for \tlab= 803.1 MeV  (\plab= 1467 MeV/c).}
\label{fig10}
\end{figure*}

The dramatic improvement that occurs when the size parameters $\alpha$ and $\beta$ take on smaller values, tends to mask the equally important
role of the final-state interaction. If the final-state interaction is turned off from the very beginning by fixing the phase parameters 
$\delta_J =0$ and $\eta_J = 1$ while $\alpha, \beta, \lambda$ are allowed to vary, $\alpha$ and $\beta$ again decrease significantly, 
which confirms an increased annihilation range within our model. Nevertheless, the resulting fit without final-state interaction is far 
from satisfactory and it is only when we include final-state interaction that we recover the fit quality discussed earlier in this section. 
This is illustrated at \tlab= 219.9 MeV (\plab= 679 MeV/c) in Fig.~\ref{fig3b}. The short-dashed curve represents the fit without
final-state interaction, and for which $\alpha = 1.27$ fm$^{-2}$ and $\beta = 1.53$ fm$^{-2}$. This short-dashed curve in Fig.~\ref{fig3b} 
should be compared with the short-dashed curve in Fig.~\ref{fig3} for which $\alpha = 2.80$ fm$^{-2}$ and $\beta = 3.23$ fm$^{-2}$. 
The differential cross section has improved significantly in the forward hemisphere. However the analyzing power is only marginally better. 
The solid line of Fig.~\ref{fig3b} is with inclusion of the final-state interaction. Similar results are obtained at the other energies, 
which confirms that in order to reproduce the LEAR data both an increased annihilation range as well as interaction in the final $\pi\pi$ 
state is necessary.

With the present parameters, we can proceed to compute other spin observables for the  \pppipi reaction. We take the 
opportunity to present for each energy considered above the predictions for the spin observables $A_{\ell s}$ and $A_{ss}$ 
introduced in Eqs.~(\ref{5}) and (\ref{6}). We remind the reader that the spin observables are related by
\begin{equation}
 A_{0n}^2 + A_{\ell s}^2 +  A_{ss}^2=1~,
\end{equation}
and that the reaction \pppipi has only three independent observables within a sign ambiguity, and of course \dsig is chosen 
as one of them. At each energy the corresponding parameters of Tables~\ref{model} and \ref{phases} are used. The results are shown in 
Figs.~\ref{fig6}--\ref{fig10}. The spin observables $A_{\ell s}$ and $A_{ss}$ exhibit again the typical structure with two extrema 
as a function of angle, which take often the form of a double dip. There is as yet no data to compare with.

\section{CONCLUSION} 

The extensive set of data of differential cross sections \dsig and analyzing powers $A_{0n}$ from the LEAR experiment \cite{hasan92}
on \pppipi in the range \plab = 360 to 1550 MeV/c can be fitted by the combined mechanisms $^3P_0$ and $^3S_1$ of the quark-antiquark
annihilation model. \\

The initial $\bar pp$ relative wave functions were taken from \cite{elbennich98}. It is important to include the final state $\pi\pi$
interaction and employ quark wave functions for proton, antiproton, and pions with radii which are slightly larger than the respective 
measured charge radii. Previously used hadron intrinsic quark wave functions \cite{bathas93} describe only the quark core of the hadrons 
without the $\bar qq$ cloud and their parameters therefore correspond to a considerably smaller radius. Increased hadronic radii lead to 
an increase in range of the annihilation mechanism and as a result amplitudes for $J=2$ and higher are much larger than before. This feature 
in the model is essential since the experimental data on $A_{0n}$ exhibit a double dip, which indicates the presence of substantial amplitudes 
of $J = 2$ and higher, already at lower momenta \plab. The relative strength of the $^3P_0$ and $^3S_1$ mechanisms shows a smooth energy 
dependence and suggests that the $^3P_0$ mechanism becomes more dominant at the higher energies. It is however noted that the pronounced 
forward and backward peaks in the cross section require the presence of both $^3P_0$ and $^3S_1$ mechanisms.

\begin{acknowledgments}
B.E. and W.M.K. thank the LPNHE and its Groupe Th\'eorie for their warm and stimulating hospitality. B.L. wishes to acknowledge 
the welcome and support of the Department of Physics and Astronomy of Rutgers University during his visits. Laboratoire de Physique 
Nucl\'eaire et de Hautes \'Energies is Unit\'e de Recherche des Universit\'es Paris 6 et Paris 7 associ\'ee au CNRS. 
This research was supported in part by the U.S. National Science Foundation Phy-9722088.
\end{acknowledgments}

\end{document}